\documentclass[
 aip,
 amsmath,
 amssymb,
 preprint,
]{ revtex4-1}

\usepackage{graphicx}
\usepackage{dcolumn}
\usepackage{bm}
\usepackage[utf8]{inputenc}
\usepackage[T1]{fontenc}
\usepackage{mathptmx}
\usepackage{etoolbox}
\usepackage[mathlines]{lineno}
\renewcommand{\selectlanguage}[1]{}

\makeatletter
\def\@email#1#2{%
 \endgroup
 \patchcmd{\titleblock@produce}
  {\frontmatter@RRAPformat}
  {\frontmatter@RRAPformat{\produce@RRAP{*#1\href{mailto:#2}{#2}}}\frontmatter@RRAPformat}
  {}{}
}%
\makeatother

\begin{document}

\title{Robust Poling and Frequency Conversion on Thin-Film Periodically Poled Lithium Tantalate}

\author{Anna Shelton}
 \email{annashelton@g.harvard.edu}
\affiliation{ 
John A. Paulson School of Engineering and Applied Sciences, Harvard University, 9 Oxford Street, Cambridge, 02138, Massachusetts, USA
}
\author{C. J. Xin}
\affiliation{
John A. Paulson School of Engineering and Applied Sciences, Harvard University, 9 Oxford Street, Cambridge, 02138, Massachusetts, USA
}
\author{Keith Powell}
\affiliation{
John A. Paulson School of Engineering and Applied Sciences, Harvard University, 9 Oxford Street, Cambridge, 02138, Massachusetts, USA
}
\author{Jiayu Yang}
\affiliation{
John A. Paulson School of Engineering and Applied Sciences, Harvard University, 9 Oxford Street, Cambridge, 02138, Massachusetts, USA
}
\author{Shengyuan Lu}
\affiliation{
John A. Paulson School of Engineering and Applied Sciences, Harvard University, 9 Oxford Street, Cambridge, 02138, Massachusetts, USA
}
\author{Neil Sinclair}
\affiliation{
John A. Paulson School of Engineering and Applied Sciences, Harvard University, 9 Oxford Street, Cambridge, 02138, Massachusetts, USA
}
\author{Marko Loncar}
\affiliation{
John A. Paulson School of Engineering and Applied Sciences, Harvard University, 9 Oxford Street, Cambridge, 02138, Massachusetts, USA
}

\date{\today}

\begin{abstract}
We explore a robust fabrication process for periodically-poled thin-film lithium tantalate (PP-TFLT) by systematically varying fabrication parameters and confirming the quality of inverted domains with second-harmonic microscopy (SHM). We find a periodic poling recipe that can be applied to both acoustic-grade and optical-grade film, electrode material, and presence of an oxide interlayer. By using a single high-voltage electrical pulse with peak voltage time of $10$~ms or less and a ramp-down time of $90$~s, rectangular poling domains are established and stabilized in the PP-TFLT. We employ our robust periodic poling process in a controllable pole-after-etch approach to produce PP-TFLT ridge waveguides with normalized second harmonic generation (SHG) conversion efficiencies of $208\%\text{W}^{-1}\text{cm}^{-2}$ from $1550$~nm to $775$~nm in line with the theoretical value of $244\%\text{W}^{-1}\text{cm}^{-2}$. This work establishes a high-performance poling process and demonstrates telecommunications band SHG for thin-film lithium tantalate, expanding the capabilities of the platform for frequency mixing applications in quantum photonics, sensing, and spectroscopy.
\end{abstract}


\maketitle

\section{Introduction}

Thin-film lithium tantalate (TFLT) has garnered increasing attention in recent years as an attractive alternative to the more mature platform of thin-film lithium niobate (TFLN) for applications in photonic integrated circuits \cite{wang_lithium_2024, zhang_ultrabroadband_2025, wang_ultrabroadband_2024, cai_high-q_2024}. Similarly to TFLN, TFLT is a ferroelectric crystal offering a wide transparency window, large electro-optic response ($\text{r}_{33}\approx30$~pm/V), and large second-order nonlinearity ($\text{d}_{33}\approx-21$~pm/V). Moreover, TFLT also features dramatically reduced birefringence \cite{zhu_integrated_2021}, enhanced DC stability \cite{yu_tunable_2024}, and, in bulk, has demonstrated a larger bandgap and a superior threshold for the photorefractive effect \cite{alexandrovski_uv_1999}, which have been notable challenges for the TFLN device community. Collectively, these properties have enabled nearly octave-spanning electro-optic frequency combs \cite{zhang_ultrabroadband_2025}, Kerr combs \cite{wang_lithium_2024}, and electro-optic modulators that support stable DC biasing on the time scale of days \cite{powell_dc-stable_2024}. In contrast, periodically poled TFLT (PP-TFLT) is only starting to emerge \cite{chen_periodic_2023, chen_continuous-wave_2025}.

Periodic poling is the periodic reversal of the domain orientation achieved in TFLN and TFLT through the application of a strong localized electric field. A number of devices have been demonstrated on periodically poled TFLN (PP-TFLN) \cite{wang_ultrahigh-efficiency_2018,lu_periodically_2019,chakkoria_ferroelectric_2025}, enabling frequency up and down conversion of interest in e.g. quantum photonics \cite{hwang_tunable_2023,saravi_lithium_2021,franken_sidewall_2024}, sensing \cite{boes_lithium_2023}, and spectroscopy \cite{wu_visible--ultraviolet_2024}. PP-TFLT offers important advantages over PP-TFLN in near-visible and shorter wavelength applications due to its lessened photorefractive effects and birefringence-induced mode mixing \cite{suntsov_periodically-poled_2024, xue_-chip_2023}. Like TFLN, the periodic poling process for TFLT is known to be sensitive to fabrication conditions, requiring precise process control to ensure uniform domain inversion and duty cycles \cite{chen_periodic_2023}. Here we explore a robust fabrication process for PP-TFLT by systematically varying fabrication parameters and confirming the quality of inverted domains. We find a periodic poling recipe that can be used for both acoustic-grade and optical-grade films, and is insensitive to electrode material and the presence of an oxide interlayer. We then employ a controllable pole-after-etch process \cite{xin_wavelength-accurate_2024} to produce PP-TFLT waveguides with normalized conversion efficiencies of $208\%\text{W}^{-1}\text{cm}^{-2}$ from telecommunications to near-visible wavelengths in line with the theoretical value of $244\%\text{W}^{-1}\text{cm}^{-2}$.

\section{Robust PP-TFLT Process}

\begin{figure*}
\includegraphics{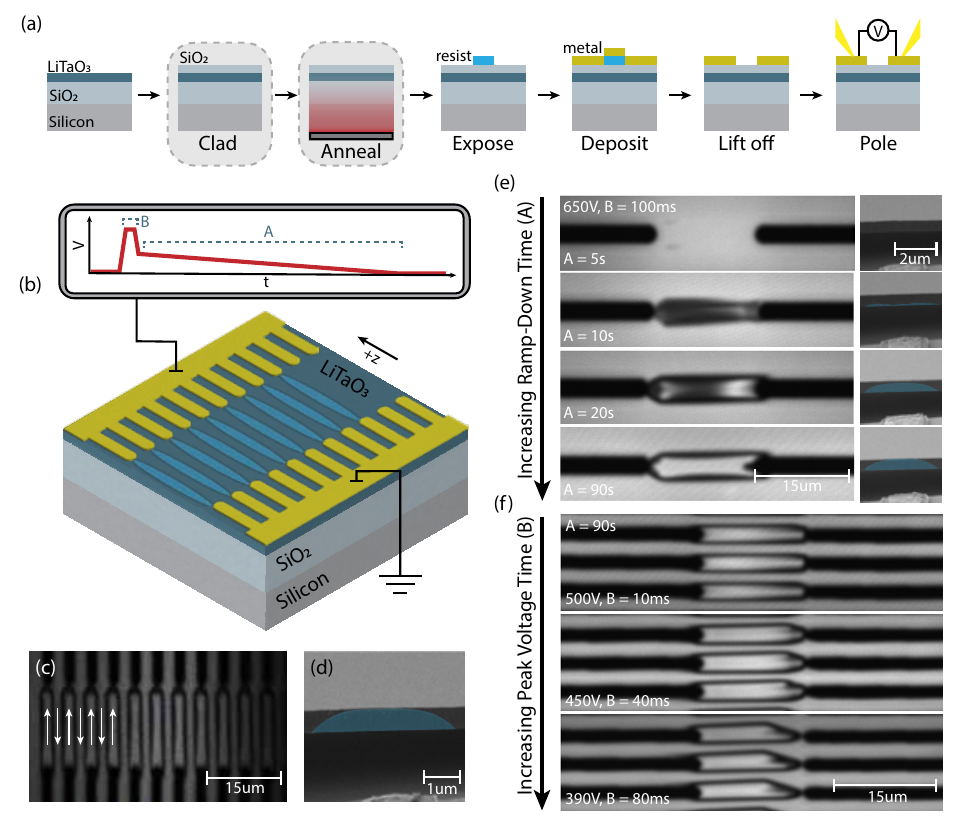}
\caption{\label{fig:fig1} (a) Fabrication flow for the robust PPLT study, with optional steps shown in grey dashed boxes. (b) Illustration of a PPLT sample during the poling process with a qualitative poling pulse form highlighting relevant pulse paramters. A is the ramp-down time, B is the peak-voltage time. (c) SHM image of a PPLT sample with a near $50\%$ duty cycle. Black rectangles at the top and bottom of the image are poling fingers. Dark lines between the fingers indicate domain walls. Bright regions indicate where the crystal domains of the TFLT are aligned. Arrows indicate the direction of the crystal domain for each bright region (d) Colorized SEM cross-sectional image of a PPLT sample after the differential etch to expose domain orientation. The poled domain is shown in color, reaching $100\%$ depth in the LT film. (e) SHM images (left) and colorized SEM cross-sectional differential etch images (right) of PPLT samples with varying pulse ramp-down times. Pulse details are noted on each SHM image. As pulse ramp down time (A) increases, poled domains penetrate deeper into the LT film. (f) SHM images of PPLT samples optimized for varying peak voltage hold times. Pulse details are noted on each SHM image. As peak voltage time (B) increases, domains become more asymmetric.} 
\end{figure*}

Our devices are fabricated on $1~\text{cm}\times1.5~\text{cm}$ TFLT chips featuring a $500$~nm thick $x$-cut LT device layer on top of $2000$~nm thick SiO$_2$ on a Si handle wafer. The fabrication starts by cleaving chips out of an acoustic-grade or optical-grade TFLT wafer (NanoLN) followed by a solvent clean. Before we perform either optical or electron-beam lithography to define poling electrodes, on some chips we deposit an oxide interlayer via plasma-enhanced chemical vapor deposition (PECVD) and anneal them in an oxygen atmosphere at $520^{\circ}$~C for 5 hours or an ambient atmosphere at $300^{\circ}$~C for 8 hours (Fig 1a). Electrodes are patterned using MicroChem LOR and Megaposit SPR220 resists in the case of optical lithography, or LOR and Zeon Chemicals ZEP520 resists in the case of electron beam lithography. Rounded-tip electrode fingers are $2~\mu$m in width and $25~\mu$m in length. Poling periods are varied from 2-5 times the poling finger width to determine what proportion of the poling period fingers should occupy to result in $50\%$ duty cycle of poled domains for maximum poling efficiency \cite{boyd_nonlinear_2008}. Electrode pads, used to land electric probes during the poling process, are approximately $100~\mu $m in width and are electrically connected to each poling finger. The high-voltage and ground electrode fingers are separated by a gap of $10-15~\mu$m. After the lithography step, an oxygen descum is performed followed by the evaporation of $50-100~$nm of a test metal (Cr, Ni, Ti, or NiCr) and $100$~nm of gold as a cap layer. Liftoff is performed, and photoresist is spun on top of the chip to prevent the electrical breakdown of air during the poling process.

\begin{figure*}
\includegraphics{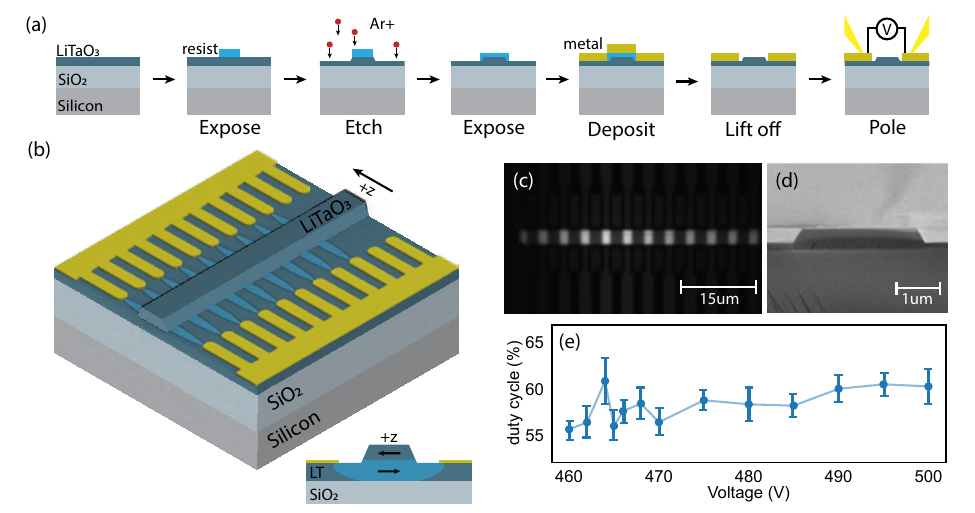}
\caption{\label{fig:fig2} (a) Fabrication flow for the PP-TFLT chip for SHG. (b) Illustration of a PPLT sample with electrodes deposited and poling performed after the ridge waveguide etching process. Cross-sectional illustration (bottom right) of a poled domain for a PPLT sample prepared as shown. The poled domain does not extend far upward into the ridge waveguide. (c) SHM image of this PPLT sample with a near $50\%$ duty cycle. (d) SEM of a PPLT waveguide facet showing a clean facet and $61^{\circ}$ sidewall angle. (e) Duty cycle as a function of peak voltage for this PPLT sample with a $5$~ms peak voltage hold time and a $90$~s ramp-down time. } 
\end{figure*}

A single pulse with the profile shown in Figure 1b is applied to each electrode. The pulse consists of $1$~kV/ms linear ramp-up rate that increases the voltage from 0 to approximately $500$~V, followed by a $100$~ms long flat-top region. This results in a poling field strength on the order of $50$~kV/mm. Finally, a linear $90$~s long voltage ramp-down is used, starting at half of the peak voltage (Fig 1b). Poled domains are then inspected under a second-harmonic microscope (SHM) to assess duty cycle and poling depth (Fig 1c). Bright regions in the SHM indicate regions where the crystal domains of the TFLT are aligned, dark regions indicate boundaries between regions of opposite crystal domain orientation, and electrodes appear as solid black shapes in the SHM. When regions between poling fingers are bright and bounded by dark boundaries, domain inversion has taken place. When the width of the domain inversion region reaches $50\%$ of the poling period, $50\%$ duty cycle is obtained. To confirm SHM images, windows are opened via optical lithography on both sides of the center line of each PP-TFLT region perpendicular to the poling fingers. We then perform an argon-based ICP-RIE etch through the full thickness of the TFLT and chemical redeposition removal process using hot KOH and SC-1. Finally a $49\%$ HF wet etch is performed, which preferentially etches the $-z$ face of lithium tantalate \cite{gao_etching_2008}, showing poled regions reach $100\%$ depth in the TFLT (Fig 1d). 

\begin{table}
\caption{\label{tab:table1} Summary of PP-TFLT devices fabricated in our study. Our material stack consists of a $500$~nm thick $x$-cut TFLT device layer on top of $2000$~nm thick SiO$_2$ on Si handle wafer (NanoLN). Devices consist of different TFLT grades (acoustic or optical), oxide interlayers, annealing treatments, lithography processes, and poling electrode contact metals. Importantly, we find the optimal poling recipe is insensitive to these variations, and all samples feature poled regionds when a poling field strength of approximately $50$~kV/mm is used. We note that higher poling field strengths are required when oxide interlayer thickness is increased and annealed an in ambient environment at $300^{\circ}$~C for 8 hours.} \begin{ruledtabular}
\begin{tabular}{cccccc}
Grade&SiO$_2$(nm)&Annealing&Lithography&Contact&kV/mm\\
\hline
Optical&-&-&EBL&NiCr&41.5\\
Acoustic&-&-&EBL&Ni&42\\
Acoustic&-&-&OL&Cr&43\\
Optical&80&520$^{\circ}$C, 5 hrs&EBL&Ni&43\\
Optical&-&-&EBL&Cr&43\\
Optical&-&-&EBL&Ni&43\\
Optical&-&-&OL&Ti&50\\
Optical&-&-&OL&Ni&50\\
Optical&-&-&EBL&Ni&50\\
Optical&50&300$^{\circ}$C, 8 hrs&EBL&Ni&53\\
Optical&100&300$^{\circ}$C, 8 hrs&OL&Ni&70
\end{tabular}
\end{ruledtabular}
\end{table}

For each combination of fabrication conditions, periodic poling was achieved with this pulse scheme (Table \ref{tab:table1}). Nearly $50\%$ duty cycle was achieved for poling periods 3 times the poling finger width, indicating a $33\%$ poling finger fill factor is ideal for reaching $50\%$ poling duty cycle for telecommunications to near-visible frequency conversion. This is similar to poling finger fill factors on PP-TFLN \cite{nagy_reducing_2019}. Additionally, shortening the ramp-down time resulted in the increasingly shallow domains (Fig 1e), with ramp-down times less than $10$~s resulting in near uniform backswitching of the TFLT between poling fingers. This pulse shape was inspired by studies on bulk PPLT \cite{meyn_fabrication_2001}, where increased time is necessary for domain stabilization to avoid domain backswitching. Poling depth was improved by increasing ramp-down time up to $90$~s. Beyond $90$~s, no further enhancement of poling depth was observed. Additionally, some devices formed triangularly-shaped domains. By shortening the pulse peak voltage time from $100$~ms to $10$~ms or shorter and reoptimizing the poling field strength, rectangular domains were recovered (Fig 1f).

\section{Second Harmonic Generation}

\begin{figure*}
\includegraphics{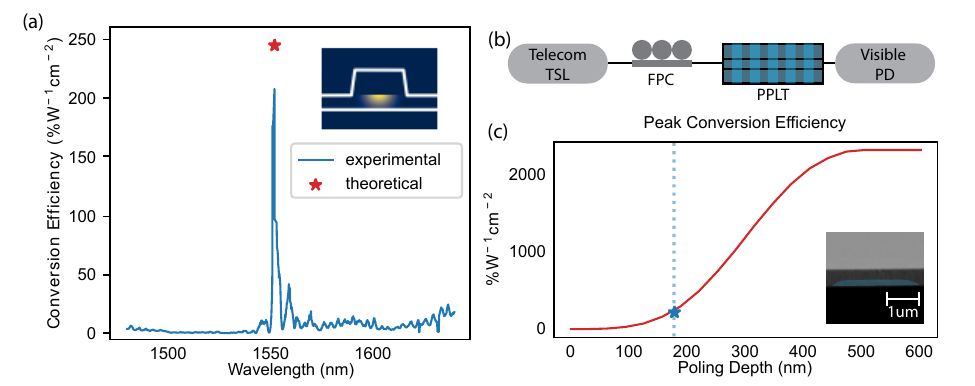}
\caption{\label{fig:fig3} (a) Normalized conversion efficiency (line) as a function of wavelength for a PPLT device, showing a maximum value of $208\%\text{W}^{-1}\text{cm}^{-2}$ at $1552$~nm. The peak theoretical normalized conversion efficiency for this device is $244\%\text{W}^{-1}\text{cm}^{-2}$ at $1552$~nm (star). (b) Illustration of the experimental setup for measuring SHG from PPLT. (c) Peak theoretical normalized conversion efficiency as a function of poled domain depth in the LT film for an identical geometry to the measured device. The poling depth of this device is highlighted by a vertical line and the experimental peak normalized conversion efficiency is plotted (star). Inset shows colorized SEM cross-sectional image of this PPLT sample after the differential etch to expose domain orientation. The poled domain is shown in color, reaching $178.7$~nm depth in the LT film.} 
\end{figure*}

We utilize a pole-after-etching approach \cite{xin_wavelength-accurate_2024} to fabricate an efficient PP-TFLT ridge waveguide on a $500$~nm thick $x$-cut optical-grade LT device layer on top of $2000$~nm thick SiO$_2$ on a Si handle, and use it for second harmonic generation (SHG) from $1550$~nm to $775$~nm. We begin fabrication by cleaving chips of size $1.8~\text{cm}\times1.8~\text{cm}$ and perform a solvent clean (Fig 2a). We perform ellipsometry on the initial film to obtain the average thickness of the device layer, then use high-resolution electron beam lithography with ma-N 2400 resist to define $2.5~\mu$m wide (top-width) and $1.4$~cm long waveguides. The chip then undergoes an argon-based ICP-RIE etch to etch approximately $300$~nm of the TLFT and chemical redeposition removal process using hot KOH and SC-1. Finally, ellipsometry is performed again to evaluate the average thickness of the remaining (un-etched) film. The etch process produces a sidewall angle of $61^{\circ}$ (Fig 2d). By fabricating the ridge waveguide before poling, actual device parameters can be used to calculate an accurate poling period and thus overcome thickness variations and fabrication imperfections. The measured film thickness, etch depth, and sidewall angle are input into simulation to extract the wavevector for the fundamental transverse electric mode at $1550$~nm ($k_{\text{fh}}$) and $775~nm$ ($k_{\text{sh}}$). The poling period ($\Lambda$) is then determined by

\begin{equation}
    \Lambda = \frac{2 \pi}{k_{\text{sh}}- 2 k_{\text{fh}}} = 3.908~\mu \text{m}.
\end{equation}

We then use the results of our robust PP-TFLT study to pole the ridge waveguide device. Poling electrode design is identical to that of the robust PP-TFLT poling study discussed above. Poling period is set to the simulated value ($3.908~\mu$m), and rounded poling finger width is set to $33\%$ of the poling period ($1.303~\mu$m), according to our study. The electrode region is set to a total length of $1$~cm to improve output second-harmonic power. Next, electron beam lithography is performed on a resist stack of LOR and ZEP520 to pattern electrodes. After patterning, an oxygen descum is performed before metal electrodes are evaporated onto the chip. Metal electrodes consist of $50$~nm Cr capped with $100$~nm Au. While we found that many metals we tested can result in good poling performance, Cr was chosen for ease of removal. Liftoff is performed and photoresist is spun on top of the chip to prevent the breakdown of air during the poling process (Fig 2b). Our robust PP-TFLT study showed that a single pulse with a peak voltage time of $10$~ms or less, poling field strength around $50$~kV/mm, and ramp-down time of $90$~s results in rectangular domains of approximately $50\%$ duty cycle. Accordingly, a single $5$~ms, $31$~kV/mm ($460$~V) pulse with $90$~s ramp-down time is applied and rectangular domains with a duty cycle of $55\%$ is found by SHM imaging (Fig 2c). The duty cycle was finely tuned by controlling the peak voltage (Fig 2e), and it approached $55\%$ as voltage was reduced to $460$~V. After poling, poling electrodes are removed by gold etchant and chrome etchant. Chips are then singulated by a fluorine- and argon-based ICP-RIE etch and Bosch etch before measurement \cite{he_low-loss_2019}.

The PP-TFLT chip is measured using an end-fire coupling setup with pump light from a continuous-wave telecommunication-wavelength tunable laser (Santed TSL-710) in the range of $1480-1640$~nm (Fig 3b). Pump light passes through a polarization controller and couples into the chip by lensed fiber. The resultant near-visible second harmonic light is collected at the output by a lensed fiber and is measured using a visible-wavelength photodetector. As the laser wavelength is swept into the designed quasi-phase matched wavelength, the second harmonic output power rapidly maximizes. The experimental conversion efficiency ($\eta_{\text{exp}}$) can be calculated

\begin{equation}
    \eta_{\text{exp}} = \frac{P_{\text{out}}}{P_{\text{in}}^2 L^2}.
\end{equation}

Where $P_{\text{out}}$ is the output second-harmonic power, $P_{\text{in}}$ is the fundamental on-chip power, and $L$ is the length of the poled region (Fig 3a). Setup and facet loss is calibrated by measuring the laser output power, fundamental output power, and second-harmonic output power and removing setup losses. After setup and facet loss calibration is performed, the maximum experimental conversion efficiency achieved on the PP-TFLT chip is $208\%\text{W}^{-1}\text{cm}^{-2}$ at $1552$~nm.

Because we use a pole-after-etch approach, poling electrodes are deposited onto the waveguide slab, and the electric field induced by the poling electrodes is insufficient to reverse the polarization direction in the uppermost portion of the ridge waveguide. To ascertain poled domain depth, windows were opened via optical lithography to bisect the ridge waveguide parallel to the central line. We then perform the same argon-based ICP-RIE etch through the full thickness of the TFLT, chemical redeposition removal process using hot KOH and SC-1, and $49\%$ HF wet etch as done in our robust poling study, revealing domains only extend through the bottom $178.7$~nm of TFLT (Fig 3c).The peak theoretical conversion efficiency ($\eta_{\text{theory}}$) can be calculated for this waveguide geometry and experimentally measured depth as 

\begin{equation}
    \eta_{\text{theory}} = \frac{2 \omega^2 d_{\text{eff}}^2} {n_{\text{fh}}^2n_{\text{sh}}\varepsilon_0 c^3} \frac{A_{\text{sh}}}{A_{\text{fh}}^2}.
\end{equation}

Where $\omega$ is the fundamental frequency, $d_{\text{eff}}$ is given by the second order nonlinear tensor, $n_{\text{fh}}$ is the refractive index of the fundamental TE mode at the fundamental frequency, $n_{\text{sh}}$ is the refractive index of the fundamental TE mode at the second harmonic, $A_{\text{fh}}$ is the fundamental TE mode area at the fundamental frequency, $A_{\text{sh}}$ is the fundamental TE mode area at the second harmonic of the, $\varepsilon_0$ is the permittivity of free space, and $c$ is the speed of light, resulting in a peak theoretical conversion efficiency of $244\%\text{W}^{-1}\text{cm}^{-2}$. This value agrees well with our experimental findings. Poling depth can be improved by e.g. poling before etching \cite{chen_continuous-wave_2025} or sidewall poling \cite{franken_sidewall_2024}. With poling depth equal to the full film thickness for the same waveguide geometry as our experimental ridge waveguides, peak conversion efficiency can reach $2314\%\text{W}^{-1}\text{cm}^{-2}$ (Fig 3c).

\section{Conclusion}

We performed a systematic study of the poling conditions of TFLT, arriving at the conclusion that the poling process is robust to both acoustic-grade and optical-grade TFLT, oxide interlayers, and poling electrode metals. For each combination tested, periodic poling with a duty cycle near $50\%$ was achieved when poling fingers were approximately $33\%$ of the poling period. A single poling pulse with a sufficiently long ramp-down duration for domain stabilization is used. We also realized a PP-TFLT ridge waveguide using a pole-after-etching process featuring a normalized conversion efficiency of $208\%\text{W}^{-1}\text{cm}^{-2}$. This value matches closely with the theoretical normalized conversion efficiency of $244\%\text{W}^{-1}\text{cm}^{-2}$. By increasing the depth of periodically poled regions in TFLT, the theoretical conversion efficiency can approach $2314\%\text{W}^{-1}\text{cm}^{-2}$ for an identical waveguide geometry. Due to its higher threshold for photorefraction, DC stability, and wider bandgap, these works pave the way for TFLT to address high power and high frequency applications in quantum optics, sensing, and spectroscopy.

\section*{Acknowledgements}
A.S. thanks Cornelis (Kees) Franken and Soumya Ghosh for discussions related to the project. A.S. thanks Meg Reardon and Ganna Savostyanova for administrative support. K.P. acknowledges support from the Harvard GRID fellowship. S.L. acknowledges support from the A*Star National Science Scholarship. 
This work was supported by Amazon Web Services (Grant No. A50791), the Air Force Office of Scientific Research (Grant No. FA9550-23-1-0333), the National Research Foundation of Korea (Grant No. NRF-2022M3K4A1094782), NASA (Grant No. 80NSSC22K0262 and 80NSSC23PB442), the National Science Foundation (Grant No. OMA-2137723, EEC-1941583, and 213806), and the Office of Navel Research (Grant No. N00014-22-C-1041).
This work was performed in part at the Harvard University Center for Nanoscale Systems (CNS); a member of the National Nanotechnology Coordinated Infrastructure Network (NNCI), which is supported by the National Science Foundation under NSF award no. ECCS-2025158.

\section*{Author Declaration}

\subsection{Conflict of Interest}
Keith Powell is co-founder of Lumina USA Inc. which is developing thin film lithium tantalate technologies.

\subsection{Author Contributions}
A.S. designed the devices, performed the experiments, and analyzed the data. A.S., C.X., and K.P. developed the fabrication process. A.S. and K.P. fabricated the devices. J.Y. and S.L. provided advice on the fabrication process and experiment. A.S., M.L., and N.S. wrote the manuscript with comments from all authors. M.L. supervised the project.

\section*{Data Availability Statement}
The data that support the findings of this study are available from the corresponding author upon reasonable request.

\bibliography{Paper2502}

\end{document}